# RePort: A Model for Remote Parental Control System using Smartphones

K.S. Kuppusamy[1], Leena Mary Francis[2], G. Aghila[3]

[1,3]Department of Computer Science, School of Engineering and Technology,
Pondicherry University, Pondicherry, India

[1]kskuppu@gmail.com, [2]rosebeauty02@gmail.com, [3]aghilaa@gmail.com

## ABSTRACT

*The mammoth growth in the information technology sector in terms of both quantity and pervasiveness has opened up a Pandora's Box of issues which are relevant in technical and social aspects. This paper attempts to address one of the critical issues among them which isconcerned with the proper utilization of computer systems and mobile devices by children and teens. This paper proposes a model termed "RePort (**Re**mote **P**arental C**o**ntrol System using Sma**rt**phones) which is aimed towards monitoring the access behaviour of computers and smartphones in general and internet to be specific, by teens and children in a private environment. The model is built such that the access behaviour shall be monitored from a remote location with the help of smartphone devices. The access- characteristics are modelled using various parameters in four different layers of the model.The proposed model is validated with a prototype implementation in the Android platform for smartphones and as a background service for computer systems. Various experiments were conducted on the prototype implementation andresults of the experiments validate the efficiency of the proposed model with respect to user's relevancy metric which is computed as 93.46%.*

## 1. INTRODUCTION

The proliferation of computer systems and smartphones among a spectrum of users has made a phenomenal impact on the day-to-day activities of billions of people across the globe. Although the benefits of information technology has reached people in all age groups, it is the teen-agers who are in the forefront of the adoption of technology [1], [2]. There are studies conducted onthe utilization of internet among the teens [3]. The result of this study shows that 93% of teens utilize the online services. 74% of population among the adults in the age group above 18 utilizes the online services.

With respect to the usage of mobile devices among the teenagers, a study reports that 78% of them uses mobile phones[4] out of which 47% of them uses smartphones.  Moreover the availability of more than 6.8 billion mobile devices across the globe emphasises the utility and the necessity of these devices across the spectrum of people.  The drastic increase in the number of mobile devices is substantiated by the fact that the quantity of devices in 2010 is 5.4 billion and 6.1 billion in 2011 [5]. With respect to the Indian context the total number of mobile phone subscription is about 906 million amounting to 73% of the population which is in line with the global average.

As with any other technology, the mobile phones do carry some adverse effects like behavioural addiction. The utility of mobile devices among the youth in specific, itself has been studied by researchers in various dimensions including the psychological analysis [1], [2].

In addition to this, the usage of social networking among the teenagers is also on a steep rise[6] which projects that 88% of parents knows that their children are using social networking tools to communicate with unknown persons. The same study reports that 67% of teenagers are aware of technique to hide their online activities from their parents. All these data emphasizes the necessity of monitoring the activities of teenagers with specific techniques.

This paper proposes a model titled "RePort" (Remote Parental Control System using Smartphones)" Theobjectives of the proposed "RePort" model are as listed below:

- Proposing a model for monitoring the usage of their computer systems and smartphonesusing remote handling techniques.
- Providing a detailed insight into the online activities and social networking tools with multitude of parameters like "time spent", "post frequency", "message sent".

The research problem addressed by this paper is defined as follows: "The prolific growth in utilization of the information technology tools like computers and smartphones by children and teenagers has put forward a serious problem of streamlining their access through parental control whose solution is attempted with the help of a multi layered remote monitoring model".

The remainder of this paper is organized as follows: In Section 2, some of the related works carried out in this domain are explored. Section 3 deals with the proposed model's mathematical representation. Section 4 is about prototype implementation and experiments. Section 5 focuses on the conclusions and future directions for this research work.

## 2. RELATED WORKS

This section walks though some of the related works which have been carried out in this domain. The proposed "RePort"model attempts to explore the online activities of the teenage users with the help of smartphones and remote monitoring techniques.

The remote accessing of a computer system using smartphones has been attempted by various researchers [7], [8]. These studies demonstrates the ability to access the computer system running an operating system like Windows from a smartphone running Android operating system [9]. Various other sensing abilities from the mobile phones have been utilized by researchers in extracting context specific details regarding the user activities [10], [11]. The problematic internet usage and its psychological impacts has been studied by various researchers [12], [13], [14].

The participation of children in online activities and the risks associated with them in the absence of parental control and with moderate control has been studied in detail in the literature [15]. The association between the cyber victimization and the behavioural modification thereafter is also analysed by a study [16]. The cyber bullying is another important issue which has been attempted by various researchers [17], [18], [19].

The parental control approaches specific to smartphones have been studied by various researchers [20], [21], [22]. These approaches adopt various filtering and monitoring techniques to observe the kids and teenagers utilization of smartphones. The tracking of their location and ability to restrict calls have also been adopted by these approaches. The parental control for teenagers differs from the kids in some aspects which has been addressed with more elaborate access control restrictions [23]. There exist various approaches to remotely access a system from smartphone devices which shall be utilized as a mechanism of parental control [24], [25], [26].

The proposed "RePort" model provides an approach with Dual parental control capabilities incorporating both the computer systems and smartphone devices. The facility to monitor the activities both with respect to the interaction with the system and online approaches has also been incorporated into the proposed "RePort" model. The details on the model are illustrated in the following section.

## 3. THE "REPORT" MODEL

This section explores about the building blocks of proposed "RePort" model for remote monitoring of system and smartphones usage by teenagers and kids in order to protect them from various cyber-threats. The block diagram of the proposed "RePort" model is as shown in Fig. 1.

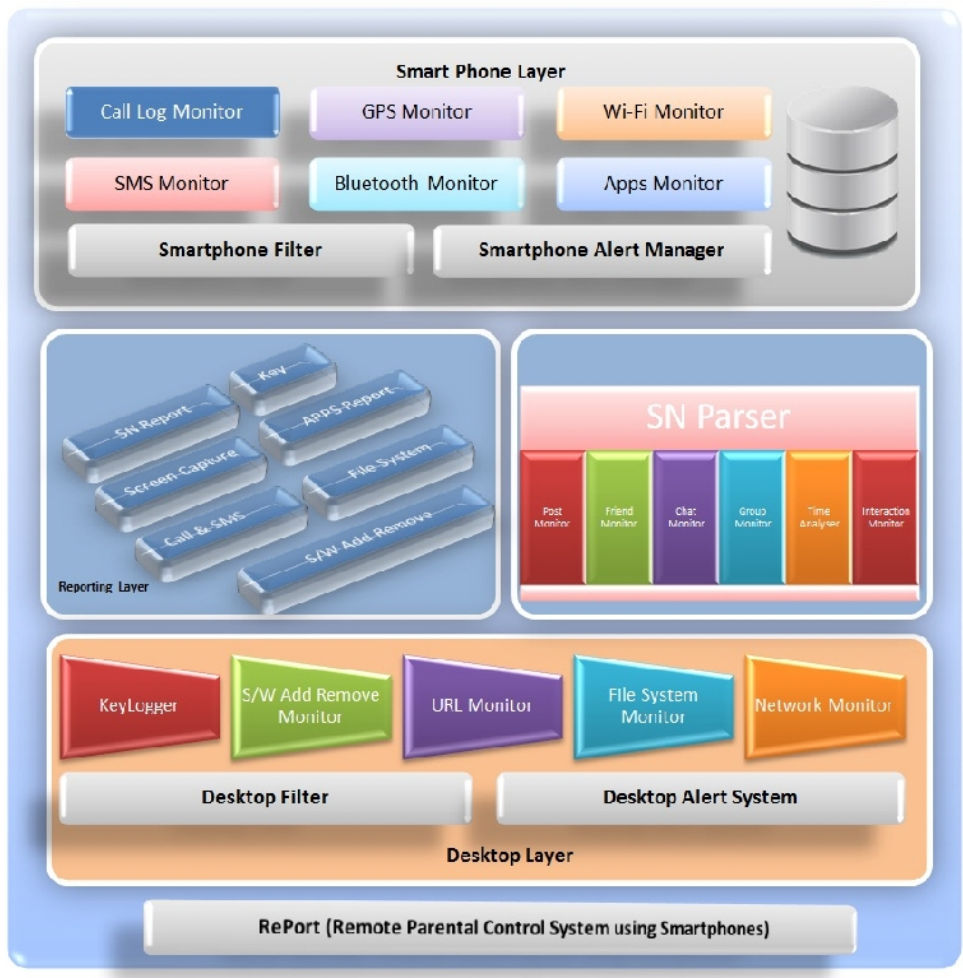

Figure 1.The "RePort" Model – Block Diagram

The proposed "RePort" model has been designed as a four layer model, as listed below:

- Smartphone Layer
- Desktop Layer
- Social Networking Layer
- Reporting Layer

Among these four layers, the Social Networking and Reporting layers receive input from both the smartphone and desktop layer.

### 3.1 Smartphone Layer

The Smartphone layer is responsible for monitoring the activities of the smartphone usage. The sub-components of the smartphone layer are as listed below:

- **Call Log Monitor:** The incoming and outgoing calls of the smartphone are tracked by this component. All these entries are stored into the local repository of the smartphone layer which would be communicated to the reporting layer at regular intervals of time. The call log monitor is as represented in (1). $\alpha_i, \alpha_j$ represents the incoming and outgoing calls respectively.

$$\Omega = \{\alpha \rightarrow [\alpha_i, \alpha_j]\} \quad (1)$$

- **SMS Monitor:** The sent and received messages from and to the smartphone are tracked by this component. As mentioned earlier, these entries are stored into local repository and communicated to the reporting layer. The SMS monitor is as represented in (2). $[\beta_i, \beta_j]$ represents the incoming and outgoing messages respectively.

$$\Omega = \begin{Bmatrix} \alpha \rightarrow [\alpha_i, \alpha_j] \\ \beta \rightarrow [\beta_i, \beta_j] \end{Bmatrix} \quad (2)$$

- **GPS Monitor:** The location of the smartphone is tracked with the help of GPS monitor. This data is utilized for tracking the movement of the teenagers. The GPS data collection is shown in (3) as $\chi \rightarrow [\forall \eta_i \in \Gamma : \chi \cup \eta_i]$.

$$\Omega = \begin{Bmatrix} \alpha \rightarrow [\alpha_i, \alpha_j] \\ \beta \rightarrow [\beta_i, \beta_j] \\ \chi \rightarrow [\forall \eta_i \in \Gamma : \chi \cup \eta_i] \end{Bmatrix} \quad (3)$$

- **Wi-Fi Monitor:** The internet data connection through the smartphone is tracked by this component. This Wi-Fi data is utilized to measure the amount of internet utilization. The WiFi monitor component's data collection is represented as $\delta \rightarrow [\forall \kappa_i \in K : \delta \cup \kappa_i]$ in (4).

$$\Omega = \begin{cases} \alpha \to [\alpha_i, \alpha_j] \\ \beta \to [\beta_i, \beta_j] \\ \chi \to [\forall \eta_i \in \Gamma : \chi \cup \eta_i] \\ \delta \to [\forall \kappa_i \in K : \delta \cup \kappa_i] \end{cases} \quad (4)$$

- **Bluetooth Monitor:** This component is responsible for tracking the files sent and received through the Bluetooth channel. This data serves as a pivotal data in tracking the external files received in the mobile phones. The Bluetooth incoming and outgoing files are indicated as $\varepsilon_i, \varepsilon_j$ respectively in (5).

$$\Omega = \begin{cases} \alpha \to [\alpha_i, \alpha_j] \\ \beta \to [\beta_i, \beta_j] \\ \chi \to [\forall \eta_i \in \Gamma : \chi \cup \eta_i] \\ \delta \to [\forall \kappa_i \in K : \delta \cup \kappa_i] \\ \varepsilon \to [\varepsilon_i, \varepsilon_j] \end{cases} \quad (5)$$

- **Apps Monitor:** The applications installed in the smartphones play an important role in the utilization of the mobile device. Though the Apps enhance the utility of the smartphone, the Apps from the unsecure sources shall lead to critical problems. In order to handle this, the installation and removal of Apps from the smartphones is monitored by this component. The apps removal and installation is represented as $\phi \to [\forall \rho_i \in P : \phi \cup \rho_i]$ in (6).

$$\Omega = \begin{cases} \alpha \to [\alpha_i, \alpha_j] \\ \beta \to [\beta_i, \beta_j] \\ \chi \to [\forall \eta_i \in \Gamma : \chi \cup \eta_i] \\ \delta \to [\forall \kappa_i \in K : \delta \cup \kappa_i] \\ \varepsilon \to [\varepsilon_i, \varepsilon_j] \\ \phi \to [\forall \rho_i \in P : \phi \cup \rho_i] \end{cases} \quad (6)$$

- **Smartphone Filter:** The smartphone filter component is responsible for detecting the improper usage of the smartphone which is carried out with the help of data collected by above said components. The "RePort" model is designed such that the utility data is constantly updated with the real-time usage data for enhancing the filtering functionality.

International Journal on Cybernetics & Informatics ( IJCI) Vol.2, No.3, June2013

- **Smartphone Alert Manager:** The smartphone filter component communicates to the smartphone alert manager component which provides alerts to the user while the improper actions are detected. The "RePort" model has provisions for customizing the levels of alert messages which would be communicated to the user.

The smartphone local repository holds the data collected by all of the monitoring components of the smartphone layer.

## 3.2 Desktop Layer

The Desktop Layer is responsible for monitoring the usage behaviour of computer systems by teenagers and children. The desktop layer has various components which are illustrated in this section. The five different monitoring components of desktop layer are represented in (7).

$$\Psi = \begin{Bmatrix} \mu \\ \nu \\ o \\ \pi \\ \varpi \end{Bmatrix} \qquad (7)$$

- **Key Logger:** The key logger component is responsible for collecting the key log data from the user. This key log shall be utilized as a source data for finding out actions performed by the user on the computer system which is represented as shown (8).

$$\Psi = \begin{Bmatrix} \mu \rightarrow [\forall \sigma_i \in \Gamma : \mu \cup \sigma_i] \\ \nu \\ o \\ \pi \\ \varpi \end{Bmatrix} \qquad (8)$$

- **Software Add Remove Monitor:** The software add remove monitor is responsible for monitoring the installation and removal of software from the system by the user. The installation of unnecessary software by the teenagers and kids shall expose them to various cyber threats. In order to avoid this, the installation and removal of software is collected as shown in (9).

$$\Psi = \begin{Bmatrix} \mu \rightarrow [\forall \sigma_i \in \Gamma : \mu \cup \sigma_i] \\ \nu \rightarrow [\forall \varsigma_i \in T : \nu \cup \varsigma_i] \\ o \\ \pi \\ \varpi \end{Bmatrix} \qquad (9)$$

- **URL Monitor:** The URL monitor performs the task of collecting all the web addresses entered by the user in browsers installed in the system. This data provides a detailed insight into the browsing behaviour of user in the system. The URL data collection is as shown in (10).

$$\Psi = \begin{Bmatrix} \mu \rightarrow [\forall \sigma_i \in \Gamma : \mu \cup \sigma_i] \\ \nu \rightarrow [\forall \varsigma_i \in T : \nu \cup \varsigma_i] \\ o \rightarrow [\forall \tau_i \in \Upsilon : o \cup \tau_i] \\ \pi \\ \varpi \end{Bmatrix} \quad (10)$$

- **File System Monitor:** The role of file system monitor is to gather the file system activities like manipulation of files and folders in the system. The process is as illustrated in (11).

$$\Psi = \begin{Bmatrix} \mu \rightarrow [\forall \sigma_i \in \Gamma : \mu \cup \sigma_i] \\ \nu \rightarrow [\forall \varsigma_i \in T : \nu \cup \varsigma_i] \\ o \rightarrow [\forall \tau_i \in \Upsilon : o \cup \tau_i] \\ \pi \rightarrow [\forall \upsilon_i \in E : \pi \cup \upsilon_i] \\ \varpi \end{Bmatrix} \quad (11)$$

- **Network Monitor:** The network monitor is responsible for gathering data related to the network activities. The network monitoring component is represented as shown in (12).

$$\Psi = \begin{Bmatrix} \mu \rightarrow [\forall \sigma_i \in \Gamma : \mu \cup \sigma_i] \\ \nu \rightarrow [\forall \varsigma_i \in T : \nu \cup \varsigma_i] \\ o \rightarrow [\forall \tau_i \in \Upsilon : o \cup \tau_i] \\ \pi \rightarrow [\forall \upsilon_i \in E : \pi \cup \upsilon_i] \\ \varpi \rightarrow [\forall \omega_i \in \partial : \varpi \cup \omega_i] \end{Bmatrix} \quad (12)$$

- **Desktop Filter:** The desktop filter component is responsible for detecting the improper usage of the computer system which is carried out with the help of data collected by above said components. The "RePort" model is designed such that the utility data is constantly updated with the real-time usage data for enhancing the filtering functionality.

- **Desktop Alert Manager:** The desktop filter component communicates to the desktop alert manager component which provides alerts to the user while the improper actions are detected. The "RePort" model has provisions for customizing the levels of alert messages which would be communicated to the user.

### 3.3 Social Networking Layer

The social networking layer monitors the activities of the user in terms of social networking functions. The proposed "RePort" model has a social networking parser with six different parameters for monitoring the activities, as shown in Figure 2.

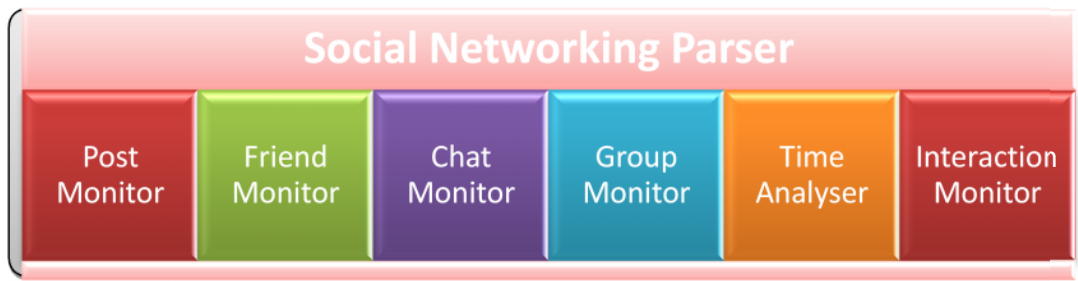

Figure 2. The Social Networking Parser

- **Post Monitor:** The post monitoring component monitors the "posts" made by the user in various social networking forums. The number of posts provides an insight into the user's intensity of social networking usage.
- **Friend Monitor:** The friend monitor analyses the friend request and response sent and received by the user.
- **Chat Monitor:** The chat monitor measure the chat activities carried out by the user. This is utilized in understanding the user's chat behaviour and intensity.
- **Group Monitor:** The group monitor is responsible for monitoring the groups in which the user has joined as a member. The improper group membership of the user leads to an alarm by the system.
- **Time Analyser:** The time analyser is responsible for measuring the overall time spent by the user in social networking. If the time goes above the threshold value set, then it would lead to an alarm.
- **Interaction Monitor:** The interaction monitor is responsible for monitoring the overall interaction of user with the social networking viz., "Like", "Share" , "+1" etc.

### 3.4 Reporting Layer

The reporting layer receives inputs from all the other three layers and it consolidates the data and provides various reports. The "Screen capture" report provides the parent with the screenshots of the monitored system taken at various intervals of time. These snapshots provide a direct view of the monitored system at regular intervals of time. The screenshot capture is triggered by the alert manager so that the screenshot shall be taken at times when there is a violation of access restriction set by the parents.

Similarly, the Apps report and Software Add/ Removal report would provide a detailed report on all the applications installed and removed both in smartphone and the system. The Key Logger data is also prepared as a report when the restricted words are detected. The call and SMS report provides the detailed report on calling and messaging.

All these reports provide the parent, the necessary details on the usage behaviour with respect to the system and smartphone by the teenagers and children.

## 4. EXPERIMENTS AND RESULT ANALYSIS

This section explores the experimentation and results associated with the proposed"RePort" model for remote monitoring the usage behaviour of system and smartphones by teenagers and children. The prototype implementation is done with the Android platform[9] for the smartphones and for the desktop the Windows operating system is chosen. The prototype application is tested with an array of hardware running Android operating system and Windows operating system. Though the prototype is made with the Android platform, the same shall be extended to other environments like iOS and Windows Mobile.

The experiments were conducted with 15 groups of users covering a spectrum from various age groups under 18. The relevance of the data reported by the "RePort" model was analysed with the inputs from adults who is responsible for monitoring the kids and teenagers.The experiments were conducted in various sessions. The results of the experiments are illustrated in Figure 3 and Table 1. During the experiments the adults were asked to mark the information rendered by the "RePort"model prototype into three different categories viz., "CRR – Completely Relevant Report", "PRR – Partially Relevant Report" and "CIR – Completely Irrelevant Report".

Table 1: RePort Information Relevance

| Session ID | CRR | PRR | CIR |
| --- | --- | --- | --- |
| 1 | 90.1 | 5.4 | 4.5 |
| 2 | 80.1 | 7.3 | 12.6 |
| 3 | 85.6 | 8.5 | 5.9 |
| 4 | 86.4 | 10.1 | 3.5 |
| 5 | 87.5 | 9.5 | 3 |
| 6 | 81.1 | 11.5 | 7.4 |
| 7 | 83.2 | 9.5 | 7.3 |
| 8 | 89.1 | 4.5 | 6.4 |
| 9 | 88.4 | 5.5 | 6.1 |
| 10 | 87.5 | 4.5 | 8 |
| 11 | 91.2 | 2.1 | 6.7 |
| 12 | 89.5 | 4.5 | 6 |
| 13 | 89.5 | 5.1 | 5.4 |
| 14 | 88.4 | 4.6 | 7 |
| 15 | 87.5 | 4.3 | 8.2 |

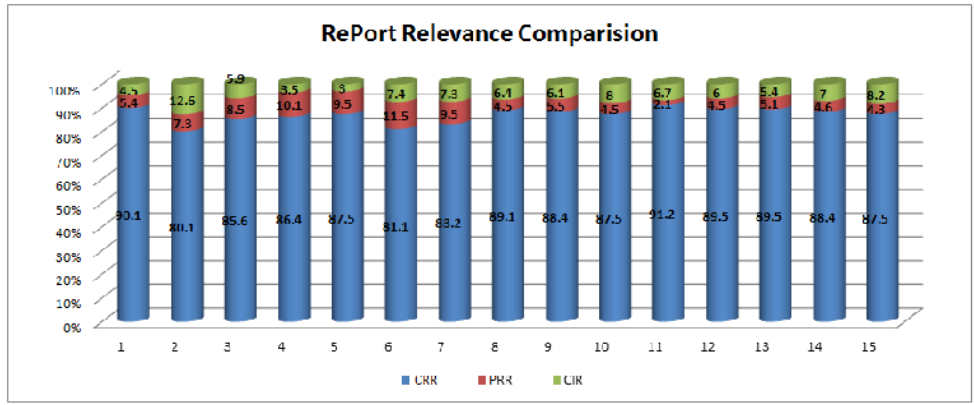

Figure 3: RePort Model Relevance Comparison

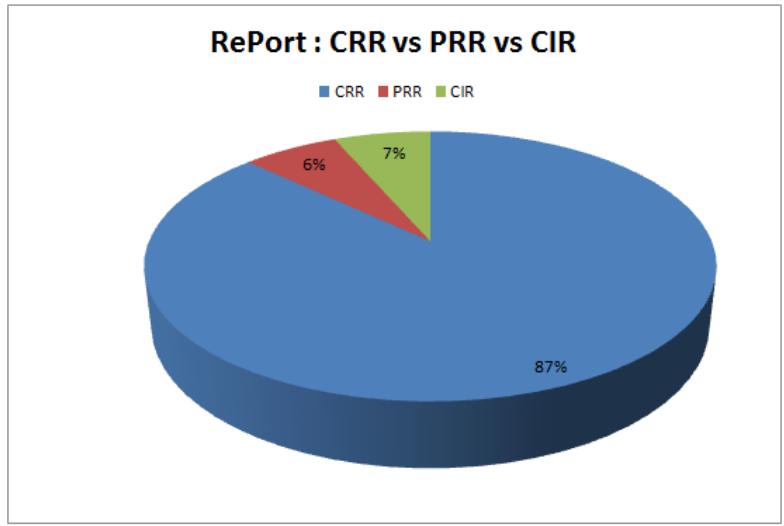

Figure 4: RePort CRR vs PRR vs CIR

The results tabulated are for a sequence of 15 sessions. It can be observed from the experimental results that the mean of CRR across the sessions is 87%, the mean of PRR as 6.46%. The CIR component is observed only at a low value of 6.53% as illustrated in Figure 4. The cumulative of CRR and PRR is evaluated as 93.46% which is the efficiency of the "RePort" model in monitoring the usage behaviour of teenagers and children with respect to smartphones and computer systems.

## 5. CONCLUSIONS AND FUTURE DIRECTIONS

The following are the conclusions derived from the "RePort" model:

- The proposed "RePort" model provides remote monitoring capability for keeping control over the usage behaviour of smartphone and computer systems by the teenagers and children.

34

- The efficiency of the "RePort" model is proven with the report relevance metric with a cumulative PRR and CRR value of 93.46%.

The future directions for the "RePort"model include the following:

- The "RePort" model shall be extended to other platforms apart from the Android and Windows operating systems.
- The "RePort" model shall be enriched with more monitoring parameters and applying specific data mining algorithms in detecting patterns in the usage behaviour.
- The model shall be further enriched by the incorporation of more reports like "Video Capture" of user's screen.

International Journal on Cybernetics & Informatics ( IJCI) Vol.2, No.3, June2013

## Authors

**Dr. K.S.Kuppusamy** is an Assistant Professor at Department of Computer Science, School of Engineering and Technology, Pondicherry University, Pondicherry, India. He has obtained his Ph.D in Computer Science and Engineering from Pondicherry Central University, and the Master degree in Computer Science and Information Technology from Madurai Kamaraj University. His research interest includes Web Search Engines, Semantic Web and Mobile Computing. He has made more than 20 peer reviewed international publications. He is in the Editorial board of three International Peer Reviewed Journals.

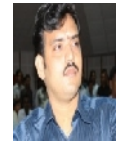

**Leena Mary Franciswas** associated with Oracle as Software Engineer before she joined as Assistant Professor at Department of Computer Science, SS College,Pondicherry, India. She has obtained her Master's degree in Computer Applications from Pondicherry University. Her area of interest includes Web 2.0 and Information retrieval.

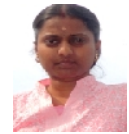

**G. Aghila** is a Professor at Department of Computer Science, School of Engineering and Technology, Pondicherry University, Pondicherry, India. She has got a total of 24 years of teaching experience. She has received her M.E (Computer Science and Engineering) and Ph.D. from Anna University, Chennai, India. She has published more than 75 research papers in web crawlers, ontology based information retrieval. She is currently a supervisor guiding 8 Ph.D. scholars. She was in receipt of Schrneiger award. She is an expert in ontology development. Her area of interest includes Intelligent Information Management, artificial intelligence, text mining and semantic web technologies.

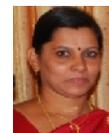